\documentclass[aps,preprint,showpacs,superscriptaddress,groupedaddress]{revtex4}  
\usepackage{graphicx}  
\usepackage{dcolumn}   
\usepackage{bm}        
\usepackage{amssymb}   

\begin{document}

\widetext

\title{Graphene Transport at High Carrier Densities using a Polymer Electrolyte Gate}
\affiliation{Department of Physics, 2 Science Drive 3, National University of Singapore, Singapore 117542, Singapore}
\affiliation{NUS Graduate School for Integrative Sciences and Engineering (NGS), Singapore 117456}
\affiliation{Department of Chemistry, 3 Science Drive 3, National University of Singapore, Singapore 117543}
\affiliation{Nanocore, 4 Engineering Drive 3, National University of Singapore 117576}
%
\author{Alexandre Pachoud} \thanks{ M. Jaiswal and A. Pachoud are equal contributors to this work.}\affiliation{Department of Physics, 2 Science Drive 3, National University of Singapore, Singapore 117542, Singapore} \affiliation{NUS Graduate School for Integrative Sciences and Engineering (NGS), Singapore 117456} 
\author{Manu Jaiswal}\thanks{ M. Jaiswal and A. Pachoud are equal contributors to this work.} \affiliation{Department of Physics, 2 Science Drive 3, National University of Singapore, Singapore 117542, Singapore} \affiliation{Department of Chemistry, 3 Science Drive 3, National University of Singapore, Singapore 117543}
\author{ Priscilla Kailian Ang} \affiliation{Department of Chemistry, 3 Science Drive 3, National University of Singapore, Singapore 117543}
\author{Kian Ping Loh}\affiliation{Department of Chemistry, 3 Science Drive 3, National University of Singapore, Singapore 117543} 
\author{ Barbaros \"{O}zyilmaz} \thanks{Corresponding Author: phyob@nus.edu.sg}\affiliation{Department of Physics, 2 Science Drive 3, National University of Singapore, Singapore 117542, Singapore} \affiliation{NUS Graduate School for Integrative Sciences and Engineering (NGS), Singapore 117456} \affiliation{Nanocore, 4 Engineering Drive 3, National University of Singapore 117576}




\date{\today}

\begin{abstract}
We report the study of graphene devices in Hall-bar geometry, gated with a polymer electrolyte. High densities of 6 $\times 10^{13}/cm^{2}$ are consistently reached, significantly higher than with conventional back-gating. The mobility follows an inverse dependence on density, which can be correlated to a dominant scattering from weak scatterers. Furthermore, our measurements show a  Bloch-Gr\"{u}neisen regime until 100 K (at 6.2 $\times10^{13}/cm^{2}$), consistent with an increase of the density. Ubiquitous in our experiments is a small upturn in resistivity around  3 $\times10^{13}/cm^{2}$, whose origin is discussed. We identify two potential causes for the upturn: the renormalization of Fermi velocity and an electrochemically-enhanced scattering rate.
\end{abstract}

\maketitle

\section{Introduction}
Since its first exfoliation from graphite in 2004 \cite{Novoselov.First}, graphene transport properties have mainly been studied in the vicinity of the Dirac point, where the dispersion relation is linear and the electrons behave as massless Dirac particles \cite{Young.Nat,Novoselov.Nat,CastroNeto.Review}. For technical reasons, the electrical properties of graphene have rarely been measured at densities beyond 10$^{13}/cm^{2}$. But the physics of graphene may well be as exciting at high charge carrier densities as it is in the vicinity of the Dirac point. As the chemical potential is shifted away from the Dirac point, the description of electrons as massless Dirac particles becomes less valid and corrections are needed to describe the physics\cite{Rotenberg.PRL,Gruneis.PRB}. Besides, recent angle resolved photoemission spectroscopy (ARPES) experiments\cite{Rotenberg.PRL} show that potassium- and calcium-doped graphene have extended van Hove singularities (VHS), a feature also present in the cuprate energy-bands and suspected by some to be responsible for their high-Tc superconducting transitions\cite{Rotenberg.PRL,Abrikosov.Physica}. However, the extended VHS of cuprates are easily accessible, whereas their graphene counterparts lie at $\sim$ 2-3 eV above the Dirac points, corresponding to electron densities greater than 2 $\times 10^{14}/cm^{2}$.

Experimentally, the realization of high carrier densities in graphene devices is limited by the requirement of thin dielectrics with high capacitance. These materials are however prone to dielectric breakdown at gate voltages required for achieving high doping.  In addition the growth and identification of graphene on various substrates remains challenging \cite{Blake.APL}. The conventional $SiO_2$ back-gate, while being suitable for identifying graphene flakes, cannot lead to carrier densities greater than 10$^{13}/cm^{2}$ in graphene. The use of high-$\kappa$ dielectrics has also been considered for achieving high doping\cite{KimSN.APL}, although this approach has been less successful. The present work uses a polymer electrolyte gate to achieve high-doping. When a potential difference is applied between two electrodes in an electrochemical cell, the ions move in the polymer matrix according to their charge polarity and accumulate to form an electric-double layer at the electrode interface. Such nanometer-size gate has a very high capacitance and can induce counter charges of equivalent density on graphene. Polymer electrolyte top-gating has been previously used to demonstrate the sensitivity of the Raman spectrum to high carrier densities in graphene \cite{Das.NatNanotech,Pinczuk.PRB}.

\section{Polymer Electrolyte Gating}
In this work, we study the electronic properties of graphene Hall devices gated with a polymer electrolyte and track the deviations from Dirac physics through density and temperature dependent transport measurements. The Hall measurements demonstrate the effectiveness of the electrolyte system in realizing high carrier densities in graphene. From transport measurements, we evaluate the relative contributions to graphene resistivity induced by different scattering mechanisms.  Monolayer graphene flakes are prepared by mechanical exfoliation on Si/$SiO_2$ substrates. The device measurements are performed on standard Hall bar and four-terminal structures fabricated with electron-beam lithography. A schematic of the device is shown in fig. 1. In addition to the electrodes on the graphene flake, Au/Cr electrodes of large surface area are also patterned within the plane of the device structure at few micron separations. While several designs for polymer electrolyte gating rely on evaporation of top gate-electrode or insertion of a Pt or Au wire in the polymer matrix\cite{Das.NatNanotech,Ozel.Nanolett}, the present design does not require positioning of the top-contact. The in-plane gate electrode can be simultaneously patterned lithographically along with the graphene contacts. Migration of metal atoms into the polymer matrix may happen in case of evaporated top contact and this contamination is also prevented. The polymer electrolyte, an aqueous dispersion of polyethylene oxide (PEO) and lithium perchlorate is then drop cast on the device and bake-dried.  

\begin{figure}
\begin{center}
\includegraphics[width=8cm]{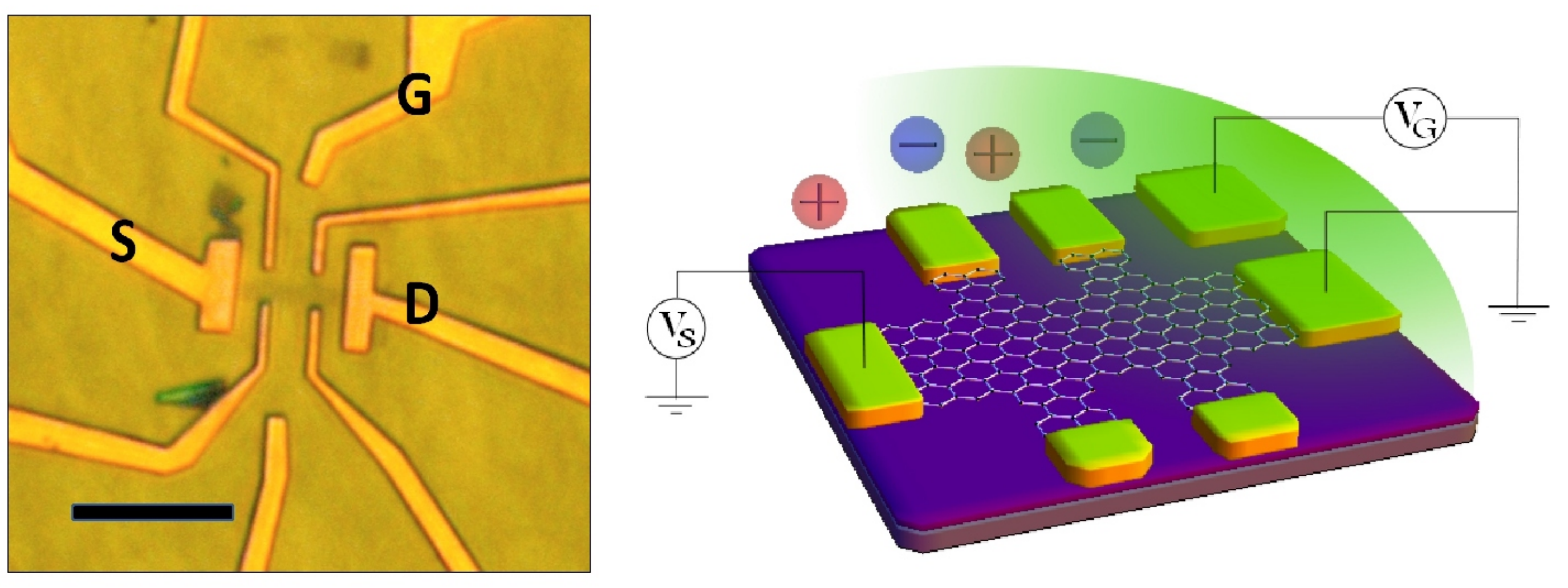}
\caption{Optical image and schematic of graphene device in Hall-bar configuration, coated with polymer electrolyte [S = Source, D = Drain, G = Polymer Electrolyte Gate]. Scale bar: 10 $\mu$m.}
\label{fig.1}
\end{center}
\end{figure}

The graphene resistance is measured at low frequencies (13 Hz) in the four-terminal configuration under low vacuum conditions. A plot showing the modulation of graphene resistivity ($\rho$) with applied polymer electrolyte gate voltage ($V_{g}$) at room temperature is shown in fig. 2(a). Due to the large interfacial capacitance arising from a nearby layer of counter ions, it is possible to obtain a large and reversible modulation in graphene resistance with the application of small voltages. The measurements are restricted to a maximum gate leakage of $\sim1$ nA. At high gate voltages (or gate leakage currents) the devices show a breakdown due to electrochemical reactions. Since the polymer is hygroscopic, the presence of adsorbed residual water contributes importantly to this leak \cite{Ueno.NatMat}, but at the same time allows a better ionic mobility. Significantly lower leakage current is observed, when the device is cooled below the ice-point of water. In addition, we note that the sweep rate of gate voltage must be slow enough to allow equilibration of the ion double layer atop graphene and measure a stabilized value of resistivity. The typical mobility of our pristine graphene samples at low doping is in the range 4000-7000 $cm^2$/Vs. Upon addition of the polymer electrolyte, the mobility of graphene remains larger than 3000 $cm^{2}$/Vs at $n \sim 10^{13}/cm^{2}$. The slope of graphene resistance $dR/dV_{g}$ (measured at half the value of maximum resistance) gated with silicon-oxide back-gate is typically 150 $\Omega V^{-1}$. This slope is enhanced significantly to ~ 3500 $\Omega V^{-1}$ when the polymer electrolyte gate is used. Upon sweeping the electrolyte gate voltage, a typical on-off ratio of 30-40 is obtained. The sharp resistance slope and high on-off ratio value are indicative of high-doping in graphene. At zero gate-voltage, graphene is found to be in a highly electron-doped low-resistance state and the charge neutrality point is shifted by -3 to -5 V. Such doping may be attributed to a higher concentration of Li$^+$ ions adsorbed in the vicinity of graphene, since the graphene has small hole-doping prior to the coating of polymer electrolyte [see fig 2(a)]. The G-band Raman peak for graphene shows a shift of 6-7 $cm^{-1}$ upon addition of the polymer electrolyte as well as a reduction in full-width at half-maximum (FWHM) [see fig. 2(b)], which further supports the electron-doping of graphene \cite{Das.NatNanotech}.

Ubiquitous in our measurements is a small upturn in resistivity observed at high gate voltages [see inset of Fig. 2(a)]. This upturn is consistently observed  in 6 graphene devices on 5 different wafers and across several sweeps for the same sample. To characterize the nature of transport at high-doping and examine the contributions to graphene resistivity, we performed Hall measurements.  The room temperature resistivity $\rho$ and Hall mobility $\mu_{H}$ of graphene devices are plotted as a function of the carrier density in fig. 3(a) and 3(b) respectively, for two devices. The mobility shows a continuous decrease between 1 $\times 10^{13}/cm^{2}$ and 6 $\times 10^{13}/cm^{2}$ , $\mu \sim 1/n$, indicating that $\rho$ approaches a saturation value. As in the low density regime ($n < 10^{13}/cm^{2}$ ), the factors determining the total resistivity include charged impurities (both from underlying substrate and from electrolyte ions), defects on the graphene lattice and phonons \cite{CastroNeto.Review,Konar.arxiv,Stauber.PRB,Chen.NatNanotech,Adam.PNAS}. While these contributions have been examined at low densities, their relative contributions at high densities can be significantly different due to high screening from carriers in graphene.  

\begin{figure}
\begin{center}
\includegraphics[width=8 cm]{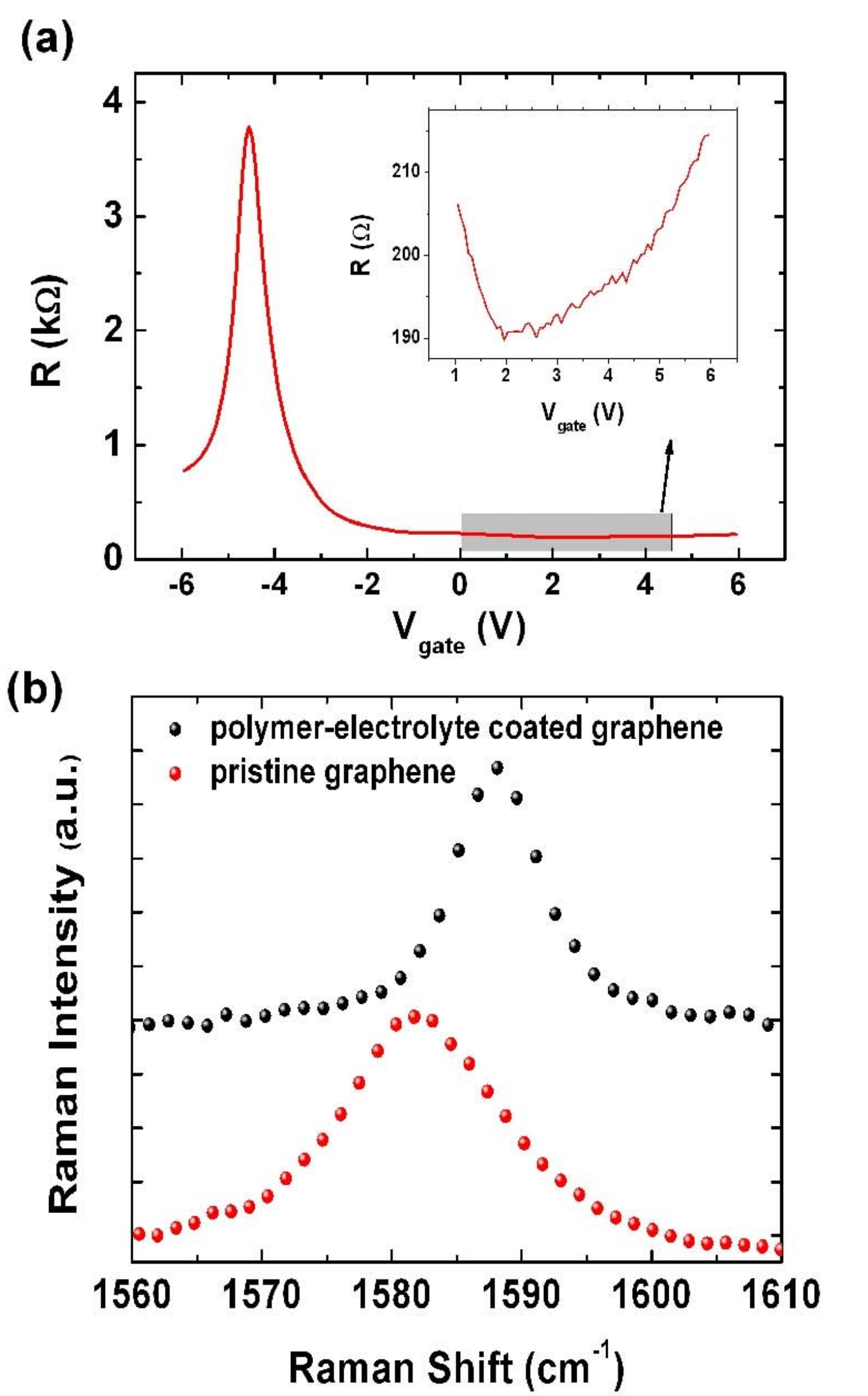}
\caption{(a) Resistance vs. polymer gate voltage for sample 1 [Inset: R vs. $V_{gate}$ in the low resistance region, showing an upturn in the device resistance] (b) G-band Raman-shift for pristine graphene (red) and polymer-electrolyte coated graphene (black). }
\label{fig.2}
\end{center}
\end{figure}

\section{Contributions to graphene resistivity}
In this experiment, graphene is sandwiched between the SiO$_2$ substrate below and the polymer layer above.  The number of electrolyte ions in the vicinity of the graphene sheet increases with the electron density, unlike the number of charged impurities at the $SiO_2$/graphene interface which remains constant. To estimate the contribution of the ions to the total resistivity of graphene, it is necessary to know their distribution in the vicinity of the graphene sheet, which is hard to obtain experimentally. Theoretically, the Poisson-Boltzmann equation is often used to describe the ion distribution in electrolyte systems\cite{Das.NatNanotech}. However, the concentration of ions estimated from this model diverges at the graphene/polymer interface while the concentration of ions is limited by the finite ionic radius, the space occupied by the polymer, and the formation of electrolyte-polymer complex. To take this into account, modified Poisson-Boltzmann equations are generally applied and/or cutoff concentrations $c_{max}$  introduced \cite{Kilic.PRE}. Following the latter approach, we modeled the ion-induced resistivity of graphene [see Appendix]. With a polymer packing density $f \le 80\%$ and an electrolyte ion effective radius around 1 nm, c$_{max}$  takes values between 10$^{25}/m^3$  and  5$\times 10^{25}/m^3$. The polymer dielectric constant is  $\epsilon$ $\sim$ 5 \cite{Das.NatNanotech}. The concentration of ions in the bulk polymer matrix is estimated to be about 5$\times10^{24}/m^3$. The gate voltage dependence of carrier density is plotted in Fig. 3(d). This can be used to experimentally estimate the total gate capacitance (polymer capacitance and quantum capacitance in series), which is of the order of $\sim1 \mu F/cm^2$ [see fig. 3(d)].  Second, we consider the influence of charged impurities from the SiO$_{2}$ substrate on the graphene resistivity. This requires an estimate of the charged impurity density $n_{imp}$ in the substrate, which can be obtained from a linear fit to the $\sigma -n$ plot at low densities for our graphene samples, prior to the addition of the polymer\cite{Adam.PNAS}. This evaluation may be an upper limit since other scatterers can also contribute to a linear density dependence of conductivity at low density \cite{Stauber.PRB}. However, by considering this upper bound, we can at least estimate the maximum contribution of charged substrate impurities to graphene resistivity.  We obtained an average value of $n_{imp} \sim 7\times 10^{11}/cm^2$ for our samples.  Therefore, calculations based on the Boltzmann theory lead to a maximum contribution of few Ohms for $n\ge 10^{13}/cm^2$. The contribution from substrate impurities is significantly lower than from electrolyte ions in the polymer matrix.

\begin{figure*}[tbp]
\begin{center}
\includegraphics[width = 16 cm]{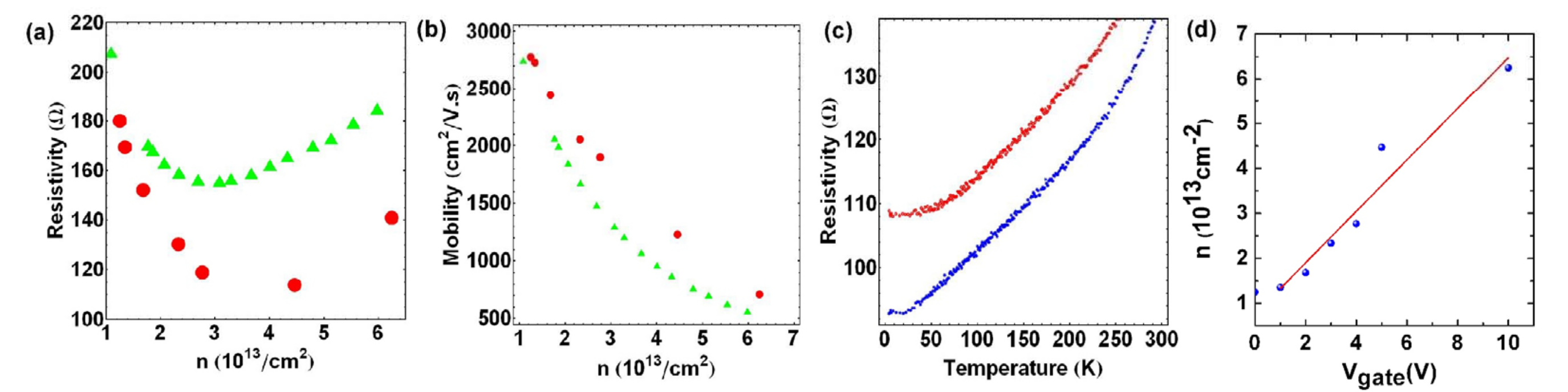}
\caption{(a) Resistivity vs. carrier concentration for Sample 1 (red) and Sample 2 (green).  (b) Hall mobility vs. carrier concentration for the same 2 samples at T = 295 K. (c) Resistivity vs. Temperature at two different densities: $n \sim$ 6.2$\times 10^{13}/cm^2$ (red), $n \sim 2.5 \times  10^{13}/cm^2$ (blue) (d) Carrier concentration vs. applied gate bias; Slope of the linear fit gives an estimate of the gate capacitance of the electrolyte gating, $C \approx 1\mu F/cm^2$. }
\label{fig.3}
\end{center}
\end{figure*}

The electrolyte ion distribution discussed above is almost temperature-independent since the ions are practically frozen below the ice-point of water. Therefore, the phonon contribution $\rho_{phonon}$ can be extracted from the temperature dependence of the graphene resistivity at high-doping. The resistivity versus temperature measurements are shown for Sample 1 down to 4 K in fig. 3(c). The upturn of resistivity observed at room temperature persists down to 4 K, since the resistivity at higher doping remains larger than the resistivity at lower doping throughout this range of temperature. The temperature-dependent part of the resistivity can be fitted by a $T^4$ law at low temperature, up to $T\sim$ 70 K and $\sim$ 100K for $n=2.5\times 10^{13}/cm^2$ and 6.2$\times 10^{13}/cm^2$ respectively. This power-law dependence can be associated to a  Bloch-Gr\"{u}neisen regime, characterized by a strong suppression of the acoustic phonon scattering rate for $T \ll T_{BG}$, where $T_{BG} = 2 \hbar v_s k_F/k_B$ is the  Bloch-Gr\"{u}neisen temperature, $v_s$ the speed of sound in graphene and $k_F$ the Fermi momentum\cite{Sarma.arxiv}. The density of phonons being governed by the Bose-Einstein law, $T_{BG}$ defines the temperature scale below which the acoustic phonon absorption rate vanishes.  Besides, the lower the temperature, the sharper the Fermi distribution and lower the acoustic phonon emission rate. These two factors lead to a complete suppression of the acoustic-phonon induced resistivity in the  Bloch-Gr\"{u}neisen regime $T\ll T_{BG}$. Further, these acoustic phonons are known to be the lowest-energy phonons graphene electrons scatter with \cite{Chen.NatNanotech}, which ensures that all phonon scattering is suppressed around 4K. Previous measurements down to 20 K and at much lower densities $n = 2\times 10^{12}/cm^2$ to  6$\times 10^{12}/cm^2$ , do not show the Bloch-Gr\"{u}neisen regime \cite{Chen.NatNanotech}. Since $T_{BG} \propto \sqrt{n}$ , the observation of a $T^4$ law to higher temperatures (up to 100 K) in our experiment is consistent with theoretical predictions. We also observe a linear regime (or non-degenerate regime) between 100 K and 170 K with a slope of  $\rho_{tot}(T)$  $\sim$ 0.13 $\Omega K^{-1}$, as the temperature becomes comparable to T$_{BG}$ (240 - 420 K), consistent with previous observations at lower densities \cite{Chen.NatNanotech}. Above 200 K, the resistivity becomes a super-linear function of the temperature, indicating that the electrons start to scatter with additional phonons, as previously discussed in the literature \cite{Konar.arxiv,Chen.NatNanotech,Morozov.PRL} for experiments at lower charge carrier densities. Finally, an estimate of   $\rho_{phonon}$ is obtained as: $\rho_{phonon} \approx \rho(295K)-\rho$(4K), which is equal to 40 to 47 $\Omega$. This almost constant phonon induced resistivity contributes to the observed $1/n$ dependence of mobility. However, even after subtracting $\rho_{phonon}$ from the total resistivity, the resulting mobility $\mu_{tot-ph} = (1/\mu_{tot}-1/\mu_{ph})^{-1}$ still shows such dependence [see fig. 4(a), 4(b)]. This indicates that other scattering mechanisms are also responsible for it, as discussed below.

\begin{figure*}[tbp]
\begin{center}
\includegraphics[width=16cm]{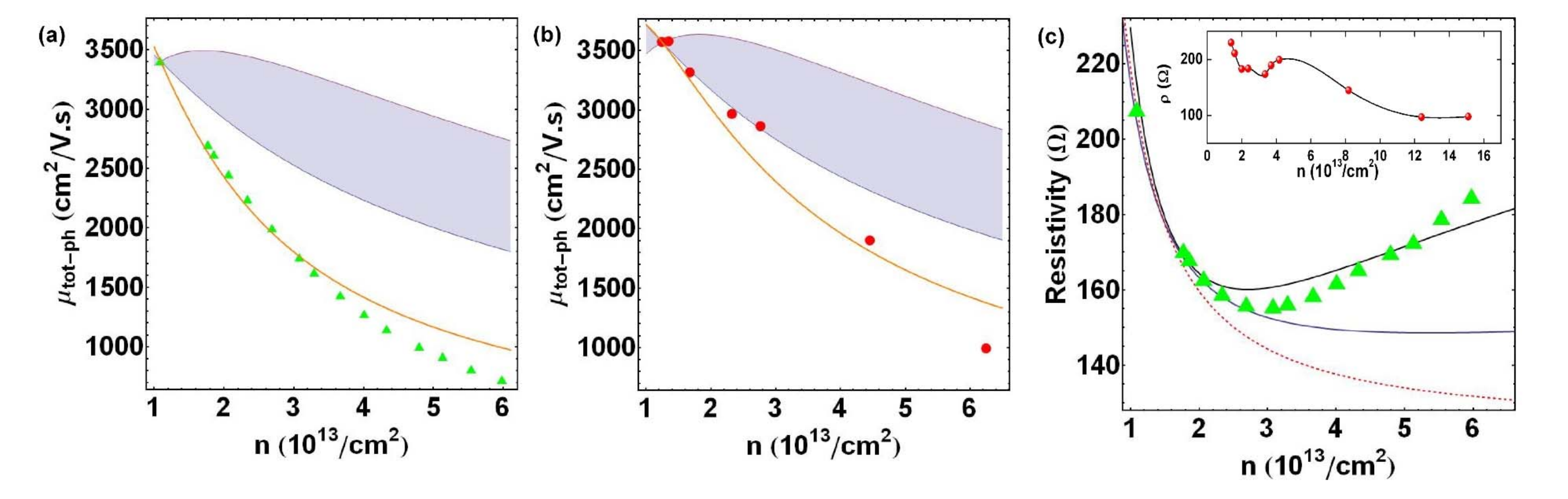}
\caption{(a) Room temperature Hall mobility  $\mu_{tot-ph}$ vs. carrier density for Sample 2. Green triangles represent experimental data, shaded blue region represents all possible $\mu_{tot-ph}$ curves that cross the first data point but do not include resistivity contribution from  weak scatterers. Orange curve is a theoretical fit after including this contribution. (b) Mobility $\mu_{tot-ph}$ vs. carrier density for Sample 1 (experimental data in red circles) (c) Resistivity vs. $n$ for Sample 2 (experimental data in green traingles). Best fits of $\rho_{tot}=\rho_{ion}+\rho_{ph}+\rho_{d}+\rho_{0}$ to resistivity data: without Fermi velocity renormalization (dashed red curve); With electron-electron interaction induced renormalization, for $e^2/(4\kappa v \hbar) \approx 0.11$ (solid blue curve); By doubling the e-e interaction coupling constant (solid black curve). This may reflect the need to include the renormalization from several other interactions as discussed in the text. [Inset: Resistivity vs. $n$ for Sample 3]}
\label{fig.4}
\end{center}
\end{figure*}

We now consider the resistivity induced by defects in the graphene lattice. Strong-potential defects such as vacancies and certain adatoms lead to a density-dependent resistivity of the form, $\rho_d = h n_d /[4e^2 n(\ln(R_0\sqrt{\pi n}))^2]$, where $R_0\approx$ 1.4 $\AA$ and $n_d$ are the size and density of these defects. Using typical values for $n_d$\cite{Stauber.PRB}, this leads to a contribution of the order of 1 $k\Omega$ near the Dirac point. This logarithmic correction leads to a sublinear defect conductivity at high density, and contributes to the decrease in mobility at high doping. However, the increase of ne$\rho_d$  with charge carrier density is too slow to reproduce the mobility behavior, as shown in fig. 4. To explain the latter, we thus consider weak scatterers inducing a \emph{constant} resistivity $\rho_0$. This scattering mechanism corresponds to a $\delta$-potential of the form, $V_{scatt}(\vec{r}$)=$V_0\delta(\vert \vec{r} \vert$) \cite{Stauber.PRB}. To estimate $\rho_0$, we fitted the theoretical expression $\mu_{tot-ph}(n) = 1/[ne(\rho_{ion}(n)+\rho_d(n)+\rho_0)]$ to the corresponding data [fig. 4(a), (b)], giving a typical value of $\sim$ 100 $\Omega$. Therefore the mobility analysis shows that the most important contributors to the resistivity of our samples at high doping are weak scatterers. 

	So far we have discussed the contributions to the graphene resistivity from ions, phonons and defects. These contributions are either nearly constant (phonons and weak scatterers) or rapidly vanishing with density (charged scatterers and strong-potential defects). It is thus surprising to consistently observe a upturn in resistivity in a finite density window near $n \sim$ 3$\times 10^{13}/cm^2$ [see fig. 3(a), 4(c)]. Note that at higher densities (1.6 $\times 10^{14}/cm^{2}$),  the resistivity decreases, then saturates, as shown for one sample (Sample 3) [see inset of fig. 4(c)]. Below, we consider possible corrections to the resistivity terms to model this observed dependence on density. At first, we note that phonons and the weak scatterers make up most of the graphene resistivity for $n > 3 \times 10^{13}/cm^2$. Therefore, it is natural to consider the upturn arising from corrections to these terms. Experimentally, the phonon contribution does not increase with density, which makes the weak scatterers the likely cause of the upturn. We therefore examine the various factors that determine the resistivity of weak scatterers, in order to identify the potential sources responsible for an increased contribution from these scatterers. Theoretically, the resistivity contribution of weak scatterers is given by\cite{Stauber.PRB,Sarma.arxiv}: 

\begin{equation}
\label{eq.1}
\rho_0(n_0,V_0,v_F) = \frac{h n_0 V_0^2}{8 e^2 [\hbar v_F(n)]^2}
\end{equation} 

where $V_{0}$ is the average impurity potential, $n_{0}$ the density of weak scatterers and $v_F(n)$ is the Fermi velocity. We examine these three factors $n_0$, $V_0$, $v_F$ for corrections to resistivity induced by weak scatterers. One possible explanation could be the electrochemically-induced creation of new defects in graphene, $\delta{n_0}$, upon application of gate volatge. However, this does not appear very plausible since the onset of upturn is seen in some samples already at small gate voltages (e.g. $V_g \sim$ 2V, $I_g < $ 50 pA for Sample 1). At low voltages, the resistivity is stable and shows negligible time dependence, precluding the formation of new defects which would be a time-dependent process. Furthermore, this resistivity increase is reversible with gate voltage and distinguishable from an irreversible increase seen at much higher gate voltages ($V_g >$ 10 V), which is likely related to electrochemical processes. A different explanation is related to the modification of the local scattering potential $V_0$ at the sites where Li$^+$ couples to the carbon lattice to forms complexes. These coupling sites may involve a finite density of pre-existing defects on the carbon lattice (\emph{e.g.} edges, vacancies), which could explain the density-dependence of resistivity after the upturn [see fig. 4(c)], as these sites get saturated with complex formation. A more interesting source for a resistivity upturn is related to the renormalization of Fermi velocity $v_F$. This renormalization has been previously shown by scanning tunneling spectroscopy for graphene on graphite and ARPES measurements on epitaxial graphene on silicon carbide substrates in our range of densities\cite{Andrei.PRL,Rotenberg.PRL}. A density-dependent renormalization of the Fermi-velocity, $v_F$ can lead to corrections to an otherwise constant  $\rho_0$.  The Fermi velocity is expected to be renormalized by direct electron-electron, Fr\"{o}hlich, electron-phonon and electron-impurity interactions \cite{CastroNeto.Review,Sarma.arxiv,Tse.arxiv,Pereira.PRB,Hu.PRB,Barlas.PRL,Calandra.PRB}. The direct electron-electron interaction is responsible for an increase of the Fermi velocity near the Dirac point, following $v \rightarrow v[1 - e^2 \ln(2a\sqrt{\pi n})/(4\kappa v \hbar)$] but this can only partially explain an increase in resistivity in our range of densities [fig. 4(c)]. Other factors such as Fr\"{o}hlich and electron-phonon interactions and disorders contribute equally to this decrease of Fermi velocity. These factors when considered together can potentially explain the magnitude of the observed increase in resistivity [see fig.4(c)] \cite{CastroNeto.Review,Sarma.arxiv,Tse.arxiv,Pereira.PRB,Hu.PRB,Barlas.PRL,Calandra.PRB}. By keeping the concentration of electrolyte ions in the compact layer $c_{max}$ around 10$^{25}/m^3$ and density of strong-potential defects $n_d$ around $10^{11}/cm^2$, it is possible to fit the $\rho -n$ curves of our samples by varying the bare defect resistivity of weak scatterers $r_0$ around 100  $\Omega$, provided that this renormalization of the Fermi velocity is taken into account. The upturn is essentially a result of the competition between a decreasing $\rho_{ion}$ and an increasing $\rho_{0}$. Therefore, the charge carrier density corresponding to the onset of the upturn increases with the ratio $c_{max}/r_0$. Besides, the decrease in Fermi velocity induced by the above-mentioned interactions, slows down at higher densities ($n\sim 10^{14}/cm^{2}$), potentially leading to a saturation of $\rho_0$ [see inset of fig. 4(c)] \cite{Tse.arxiv,Hu.PRB,Mahan.Book,Attacalite.PSS,Calandra.PRB}. It also follows that for samples with a large enough $c_{max}/r_0$, the upturn is expected to be suppressed. Finally, note that the Fermi velocity dependence of the phonon-induced resistivity and the renormalization of the former do not contradict the fact that the $\rho -T$ curves [see fig.3(c)] remain almost parallel in the linear regime. Due to the screening of the Coulomb interaction between carbon atoms, the sound velocity and the deformation potential decrease with charge carrier density. This limits the influence of a decrease of Fermi velocity on $\rho (T)$. Before we conclude, note that at low densities ($n < 5 \times 10^{12}/cm^2$), the graphene resistivity is already strongly density-dependent from Coulomb and strong potential defect scattering. Thus any corrections arising from density-dependent renormalization of Fermi velocity in the low density regime are hard to discern in transport measurements. 

\section{Concluding Remarks}
In summary, we have demonstrated high electron densities (6.2$\times 10^{13}/cm^2$) in graphene with a polymer electrolyte gate. A Bloch-Gr\"{u}neisen regime was observed between 4 K and 100 K, a clear sign of large Fermi temperatures. The density-dependence of the mobility and resistivity of our samples were analyzed by considering various scattering mechanisms: Coulomb scattering from the electrolyte ions, electron-phonon scattering, and electron-impurity scattering. Vacancies, cracks and certain adatoms are important scatterers in the low density regime. However, weak scatterers are the most important scatterers in our range of densities  ($n > 10^{13}/cm^2$), as suggested by the $1/n$ density dependence of mobility.  The resistivity versus carrier density graphs, obtained from Hall measurements, show an upturn for densities around 3$\times 10^{13}/cm^2$, and possible corrections to resistivity of weak scatterers are discussed. While the devices reported in this paper allowed to reach electron densities significantly higher than obtained with conventional dielectrics, further improvements are needed to explore the physics of graphene in the vicinity of the van Hove singularity.
\section{Appendix}
The resistivity induced by the electrolyte ions of the polymer is modeled within the framework of Boltzmann theory. Away from the Dirac point, the resistivity is given by $\rho$ = $2/[e^2 v_F^2(n) D(n) \tau (n)]$ where $v_F$ is the Fermi velocity, $D(n)$ the density of states, and $\tau$ the scattering relaxation time. We compute the scattering rate induced by a 2D layer of charged particles situated at a distance $z$ above the graphene plane by using a screened scattering potential, $V_c(q,z) = [2 \pi e^2 e^{-qz}]/[\kappa(q+q_{TF})]$. In a classic Poisson-Boltzmann approach where the steric effects are not taken into account, the concentration of ions diverges in the vicinity of the graphene sheet. However, as the gate voltage is applied, charged ions form a compact layer of thickness $\lambda$$_c$ and concentration $c_{max}$ atop graphene\cite{Kilic.PRE}. The total scattering rate $\tau^{-1}$ is obtained by integrating over the compact layer and the corresponding resistivity reads $\rho_{ion}(n) =[\pi \hbar c_{max}I_{ion}]/[8e^2 (\pi n)^{3/2}]$, where $I_{ion} = \int^1_0 \frac{u \sqrt{1-u^2}}{{(1+2u/q_s)}^2}(1-e^{-4 \lambda_c k_F u})\, du$ $\approx$ 0.13, since the exponential factor is small for typical values of compact layer thickness ($\lambda_c \sim$ 10 nm) and graphene carrier densities ($k_F \approx 1$ $ nm^{-1}$) in our experiment. Note $q_s = q_{TF}/k_F$ in the integral.

\acknowledgments
The authors thank the support of Singapore National Research Foundation under NRF-CRP grant R-143-000-360-281, NRF RF award no. NRFRF 2008-07, NUS SMF Award, US Office of Naval Research (ONR and ONR Global), and by NUS NanoCore.

\end{document}